\documentclass[aps,prd,reprint,groupedaddress, showpacs, nofootinbib]{revtex4-1}
\usepackage{graphics,graphicx,epsfig}
\usepackage{amsmath, color}
\usepackage[subnum]{cases}
 
\begin{document}

\title{Testing typicality in multiverse cosmology}
\author{Feraz Azhar}
\email[Email address: ]{feraz.azhar@alumni.physics.ucsb.edu}
\affiliation{Department of History and Philosophy of Science, University of Cambridge, Free School Lane, Cambridge, CB2 3RH, United Kingdom}

\date{\today}

\begin{abstract}
In extracting predictions from theories that describe a multiverse, we face the difficulty that we must assess probability distributions over possible observations, prescribed not just by an underlying theory, but by a theory together with a conditionalization scheme that allows for (anthropic) selection effects. This means we usually need to compare distributions that are consistent with a broad range of possible observations, with actual experimental data. One controversial means of making this comparison is by invoking the `principle of mediocrity': that is, the principle that we are typical of the reference class implicit in the conjunction of the theory and the conditionalization scheme. In this paper, I quantitatively assess the principle of mediocrity in a range of cosmological settings, employing `xerographic distributions' to impose a variety of assumptions regarding typicality. I find that for a fixed theory, the assumption that we are typical gives rise to higher likelihoods for our observations. If, however, one allows both the underlying theory and the assumption of typicality to vary, then the assumption of typicality does not always provide the highest likelihoods. Interpreted from a Bayesian perspective, these results support the claim that when one has the freedom to consider different combinations of theories and xerographic distributions (or different `frameworks'), one should favor the framework that has the highest posterior probability; and then from this framework one can \emph{infer}, in particular, how typical we are. In this way, the invocation of the principle of mediocrity is more questionable than has been recently claimed.
\end{abstract}

\maketitle

\section{Introduction}\label{SEC:Introduction}

A generic prediction of the theories that underpin our current understanding of the large-scale structure of the universe, is that the observable universe is not all that exists, and that we may be part of a vast landscape of (as yet) unobserved domains where the fundamental constants of nature, and perhaps the effective laws of physics more generally, vary. The predominant approach to characterizing this variability rests on theory-generated probability distributions that describe the statistics of constants associated with the standard models of particle physics and cosmology. The hope remains that plausible descriptions of such multi-domain universes (henceforth `multiverses'), generated, for example, from inflationary cosmology~\cite{vilenkin_83, linde_83, linde_86} or the string theory landscape~\cite{bousso+polchinski_00, kachru+al_03, freivogel+al_06, susskind_07}, will yield prescriptions for calculating these distributions in unambiguous ways. Subsequent comparisons with our observations would allow us to ascertain which multiverse models are indeed favored. 

To be more precise, one expects theories that describe a multiverse to set down a likelihood for observations we might make, given both the theory under consideration, and conditions that restrict the vast array of domains to ones in which we might arise. This latter conditionalization is naturally couched in terms of conditions necessary for the existence of `us', as defined by relevant features of the theory. The need for such `anthropic' conditionalization, as captured, for example, in what has become known as Carter's `Weak Anthropic Principle'~\cite{carter_74}, is predicated on the presumption that most of the domains described by theories of the multiverse will not give rise to the specialized structures we see around us, nor indeed to complex biological life~\cite{hartle_07}. Under this scenario any observation we might make conditionalized on theory alone, would prove to be unlikely; and one should therefore restrict one's attention to relevant domains so as to secure relevant probabilities for possible observations. 

An appropriate conditionalization scheme might make our observations more likely: but how likely should they be, before we can count them as having been successfully predicted by the conjunction of a theory and a conditionalization scheme? One proposed solution to this problem is known as the `principle of mediocrity'~\cite{vilenkin_95}: in more current terminology, it assumes that we should reason as though we are typical members of a suitable reference class (see also~\cite{gott_93, page_96, bostrom_02}). Under this assumption, for appropriately conditionalized distributions, as long as our observations are within some `typical' range according to the distribution, we can count them as being successfully predicted. The assessment of this assumption is the topic of this paper, and constitutes a central concern in extracting predictions from any theory of the multiverse.

The principle of mediocrity, or the `assumption of typicality'---as it will also be referred to in this paper---is not without its critics~\cite{weinstein_06, smolin_07, hartle+srednicki_07}. A key issue involves how one defines the reference class with respect to which we are typical~\cite{garriga+vilenkin_08}. This problem is even more stark given our ignorance of who or what we are trying to characterize, and the precise physical constraints we need to implement in order to do so~\cite{weinstein_06, azhar_14}. 

Rather than assessing this principle from a primarily conceptual point of view, we propose to test it \emph{quantitatively}. In particular, we investigate how well it does in terms of accounting for our data in comparison to other assumptions regarding typicality, in a restricted set of multiverse cosmological scenarios. We do this by extending the program of~\citet{srednicki+hartle_10} to explore a variety of assumptions regarding typicality, building these assumptions into likelihoods for possible observations through the use of `xerographic distributions' (in the terminology of~\citet{srednicki+hartle_10}). The goal then is to find the conjunction of a theory and a xerographic distribution (which they call a `framework') that gives rise to the highest likelihoods for our data. 

I will show that (1) for a fixed theory, the assumption that we are typical gives rise to higher likelihoods for our observations; but (2) if one allows both the underlying theory and the assumption of typicality to vary, then the assumption of typicality \emph{does not} always provide the highest likelihoods. Interpreted from a Bayesian perspective, these results provide support for the claim that one should try to identify the framework with the highest posterior probability; and then from this framework, one can \emph{infer} how typical we are.

The structure of this paper is as follows. In section~\ref{SEC:Xerographic_Distributions}, I outline the general formalism within which I will be investigating assumptions regarding typicality, including the introduction of a statement of the principle of mediocrity adapted to our specific purposes. Section~\ref{SEC:Multiverse_Model} introduces the multiverse model we will analyze (which is a generalization of the cosmological model of~\citet{srednicki+hartle_10}), derives the central equations for relevant likelihoods from which we will eventually test assumptions regarding typicality, and shows that these likelihoods reduce to the results of~\citet{srednicki+hartle_10} under the appropriate simplifying assumptions. Explicit tests of the principle of mediocrity are presented in section~\ref{SEC:Results}, and we conclude in section~\ref{SEC:Discussion}  with a discussion of the context in which one should interpret the results of these tests. So I turn first to a description of the general formalism within which I will be working.

\section{Xerographic Distributions}\label{SEC:Xerographic_Distributions}

\subsection{Generalities}\label{SEC:Generalities}

I begin by outlining the formalism of~\citet{srednicki+hartle_10}, recasting relevant parts of their discussion to suit our computations in the next section. 

In general multiverse scenarios, it is possible that any reference class of which we believe we are a member, may have multiple members. Indeed, it is plausible that our accumulated data $D_{0}$, which gives a detailed description of our physical surroundings, might be replicated at different spacetime locations in the multiverse. A theory $\mathcal{T}$ describing this multiverse scenario, will, in principle, generate a likelihood for this data which we will denote by $P(D_{0}|\mathcal{T})$. This corresponds to a `third-person' likelihood in the terminology of~\citet{srednicki+hartle_10}---that is, a likelihood that does not include any information about which member of our reference class we might be. The quantity that takes this \emph{indexical} information into account is a `first-person' likelihood and will be denoted by $P^{(1p)}(D_{0}|\mathcal{T}, \xi)$, in accordance with the notation of~\citet{srednicki+hartle_10}. The added ingredient here is the \emph{xerographic distribution} $\xi$, a probability distribution that we specify \emph{by assumption}, that encodes our belief about which member of our reference class we happen to be. Its functional form is independent of a given theory $\mathcal{T}$, and together with such a theory, constitutes a `framework' $(\mathcal{T}, \xi)$ (in the notation of~\cite{srednicki+hartle_10}). Thus the transition from a third-person likelihood $P(D_{0}|\mathcal{T})$ to a first-person likelihood $P^{(1p)}(D_{0}|\mathcal{T}, \xi)$ is effected by two ideas: (i) the conditionalization scheme, which (as mentioned in section~\ref{SEC:Introduction}) specifies our reference class, and (ii) a probability distribution over members of our reference class.

In the case where there exist $L$ members of our reference class at spacetime locations $x_{l}$ for $l = 1,2,\dots,L$, we let the probability that we are the member at location $x_{l}$ be denoted by $\xi_{l}$. So the xerographic distribution is just the sequence of probabilities $\xi := \{\xi_{l}\}_{l=1}^{L}$, and will always be chosen so that it is normalized to unity: $\sum_{l=1}^{L}\xi_{l}=1$. We will assume throughout this paper that the total number of members $L$ is finite. 

The assumption that we are a typical member of this reference class, is then the statement that the probability that we are any one of these members is the same, and thus the xerographic distribution is given by the uniform distribution: $\xi_{l} = \frac{1}{L}$. Correspondingly, the assumption that we are atypical of this reference class will be given by xerographic distributions that deviate from the uniform distribution. 

How then do we propose to compute the first-person likelihood $P^{(1p)}(D_{0}|\mathcal{T}, \xi)$? To do so, we will introduce a few conventions that will form the basis of the general discussion here, and also the basis of the more specific examples we will pursue in the following two sections.  

Assume then that there exists a finite set of $N$ distinct domains in a multiverse, within each of which `observers' may exist with some probability that may depend on the particular domain being considered. We will only track the existence or non-existence of observers in each domain, without concerning ourselves with issues such as: whether or not these observers are anything like `us'\footnote{The simplicity of the scenarios we will consider makes this less egregious an assumption than it would otherwise be. Indeed, as mentioned in section~\ref{SEC:Introduction}, definitions of observers and the subsequent specification of appropriate reference classes is a thorny issue, but we will not need to engage with it here (see~\cite{hartle+srednicki_07} for further discussion).}, precisely where in these domains these observers might be located, and how many observers might exist in a domain. We thereby consider only whether a domain has observers in it or does not. Thus, there exists a total of $2^{N}$ possible configurations of observers in domains across the entire multiverse. We will denote each such configuration by an $N$-dimensional vector $\vec{\sigma}$ of binary digits, where $\sigma_{i} = 1$ denotes the existence of observers in domain $i$, and $\sigma_{i} = 0$ denotes there are no observers in that domain. The set of all such configurations $\vec{\sigma}$ will be denoted by $\mathcal{K}$, and we will denote the probability of a configuration $\vec{\sigma}$ by  $P(\vec{\sigma})$.

We live inside one of these domains and observe some data. Let $D_{0}$ denote the data that there exist observers who see this same data. We will take a theory $\mathcal{T}$ to describe an \emph{observable} fact about each domain: in the model introduced in section~\ref{SEC:Multiverse_Model}, this will be the value of a binary quantity. So the probabilities $P(\vec{\sigma})$ will be given in general, independently of $\mathcal{T}$. But $\mathcal{T}$ will determine the subset $\mathcal{K}_{D_{0}}({\mathcal{T}})$ of those configurations $\vec{\sigma}$ in which $D_{0}$ is observed ($\mathcal{K}_{D_{0}}({\mathcal{T}})\subset\mathcal{K}$). The sum of the probabilities of the configurations $\vec{\sigma}$ belonging to the subset $\mathcal{K}_{D_{0}}({\mathcal{T}})$, is just the theory-generated third-person likelihood for our data: $P(D_{0}|\mathcal{T}) = \sum_{\vec{\sigma}\in\mathcal{K}_{D_{0}}(\mathcal{T})}P(\vec{\sigma})$. This specifies the likelihood that there is at least one observing system in the multiverse that witnesses the data $D_{0}$. 

To derive a first-person likelihood, we need to assume a xerographic distribution by firstly specifying a reference class that could plausibly describe `us'. Two natural reference classes to consider (which we will further develop in section~\ref{SEC:Multiverse_Model}), following~\cite{srednicki+hartle_10}, are $(i)$ the reference class of all observers who witness our data $D_{0}$, and $(ii)$ the reference class of all observers (who do not necessarily see our data $D_{0}$). In either case, for any particular possible observer configuration $\vec{\sigma}$, the xerographic distribution encodes the probability that we are the reference-class member at some specified location. Owing to the simplicity of our model, a `location' will correspond to a multiverse domain. In general, for a given $\mathcal{T}$ and given $\vec{\sigma}\in\mathcal{K}_{D_{0}}(\mathcal{T})$, only a subset of locations $L_{D_{0}}(\vec{\sigma},\mathcal{T})$, will contain observers who see our data $D_{0}$. From these considerations, we can write down the appropriate first-person likelihood $P^{(1p)}(D_{0}|\mathcal{T}, \xi)$ as follows:
\begin{equation}\label{EQN:1stPerson}
P^{(1p)}(D_{0}|\mathcal{T}, \xi) := \sum_{\vec{\sigma}\in\mathcal{K}_{D_{0}}(\mathcal{T})} P(\vec{\sigma})  \sum_{l'\in L_{D_{0}}(\vec{\sigma},\mathcal{T})}\xi_{l'}\;.
\end{equation}
To be clear, \emph{for each configuration} $\vec{\sigma}$ in the above sum over configurations, one needs to compute the appropriately normalized xerographic distribution, subsequently summing that distribution over only those locations that could indeed correspond to us.\footnote{The notation adopted in Eq.~(\ref{EQN:1stPerson}) is not to be confused with the claim that the theory in any way \emph{determines} the xerographic distribution---it does not. We are simply spelling out how our assumptions regarding which possible member of a suitable reference class we might be, enter into the determination of first-person likelihoods. This will become clearer when we consider concrete cosmological scenarios below---see Eq.~(\ref{EQN:1stPersonO}) for a preview.}

We note that for the reference class of all observers who witness our data $D_{0}$ (called $(i)$ above): for each theory $\mathcal{T}$ and each configuration $\vec{\sigma}\in\mathcal{K}_{D_{0}}(\mathcal{T})$, the subset of locations that contain observers who witness our data $D_{0}$, namely $L_{D_{0}}(\vec{\sigma},\mathcal{T})$, is the entirety of the set of locations over which the xerographic distribution can be nonzero. Thus, by the normalization condition the xerographic distribution satisfies, we have, in Eq.~(\ref{EQN:1stPerson}):  $\sum_{l'\in L_{D_{0}}(\vec{\sigma},\mathcal{T})}\xi_{l'} = 1$. For this reference class therefore, the first-person likelihood is independent of the functional form of the xerographic distribution, and reduces to the appropriate third-person likelihood: $P^{(1p)}(D_{0}|\mathcal{T}, \xi) \longrightarrow P(D_{0}|\mathcal{T})$.

\subsection{Preferred xerographic distributions}\label{SEC:Bayesian}

The likelihood introduced in Eq.~(\ref{EQN:1stPerson}) can be analyzed from a Bayesian perspective, which, under the appropriate conditions, allows us to pick out a preferred xerographic distribution. As detailed by~\citet{srednicki+hartle_10}, one can compute the posterior probability $P^{(1p)}(\mathcal{T}, \xi|D_{0})$ by Bayes' theorem
\begin{equation}
P^{(1p)}(\mathcal{T}, \xi | D_{0}) = \frac{P^{(1p)}(D_{0}|\mathcal{T}, \xi) P(\mathcal{T}, \xi)}{\sum_{(\mathcal{T}, \xi)}P^{(1p)}(D_{0}|\mathcal{T}, \xi) P(\mathcal{T}, \xi)},
\end{equation}
where $P(\mathcal{T}, \xi)$ is the prior probability of the framework $(\mathcal{T}, \xi)$. We will be working below in a simplified setting where we assume we have a set of equally plausible frameworks at our disposal. That is, each framework enters into our Bayesian analysis with an equal prior. This implies the posterior probability is simply proportional to the likelihood:
\begin{equation}
P^{(1p)}(\mathcal{T}, \xi|D_{0}) \propto P^{(1p)}(D_{0}|\mathcal{T}, \xi),
\end{equation}
and the question of which framework is to be preferred (by virtue of having the highest posterior probability), becomes a question of which framework gives rise to the highest likelihood $P^{(1p)}(D_{0}|\mathcal{T}, \xi)$. It is from this preferred framework that one can select a preferred xerographic distribution. Given that we will use these distributions to encode assumptions regarding typicality, this selection will allow us to infer, in particular, how typical we are. 

\subsection{Testing mediocrity}\label{SEC:Testing_typicality}

How then do we propose to test the principle of mediocrity? In the language introduced above, we can formulate a more precise version of the principle of mediocrity as follows:
\begin{quote}
{\bf PM}: We are typical of the entirety of the reference class of observers in the multiverse who measure our data $D_{0}$. That is, in cases where there are (a finite number of) $L$ observers who measure $D_{0}$, situated at spacetime locations $x_{l}$ for $l = 1,2,\dots, L$, the probability that we are any one of these observers is $\frac{1}{L}$. The corresponding xerographic distribution is given by $\xi_{l} =\frac{1}{L}$, for $l = 1, 2, \dots, L$. 
\end{quote}
In what follows, we will denote the xerographic distribution implementing {\bf PM} by $\xi^{\textrm{PM}}$. Any other xerographic distribution, with either a non-uniform distribution over the reference class referred to in {\bf PM}, or else any distribution over a reference class that is not the one referred to there, constitutes some form of \emph{non-mediocrity}.

With the assumptions of section~\ref{SEC:Bayesian} in mind, we propose to test the principle of mediocrity by comparing likelihoods $P^{(1p)}(D_{0}|\mathcal{T}, \xi^{\textrm{PM}})$ against $P^{(1p)}(D_{0}|\mathcal{T}^{\star}, \xi^{\star})$, where $\xi^{\star}\neq\xi^{\textrm{PM}}$, and where we allow for the possibility that the underlying theory $\mathcal{T}$ can vary as well.

\section{Extending the cosmological model of Srednicki and Hartle}\label{SEC:Multiverse_Model}

To explicate the schema of section~\ref{SEC:Xerographic_Distributions}, and to probe the plausibility of {\bf PM} in more concrete settings, we now construct a generalization of the cosmological toy model presented by~\citet{srednicki+hartle_10} (see also their earlier paper~\cite{hartle+srednicki_07}). In section~\ref{SEC:SH_multiverse}, we will demonstrate that our results for likelihoods for this extended model reproduce theirs under the appropriate simplifying assumptions. 

\subsection{Model preliminaries}\label{SEC:Model_preliminaries}

Let $\mathcal{V} = \{1,2,\dots,N\}$ label $N$ distinct domains in a multiverse, each of which is assumed to have one of two `colors', \emph{red} or \emph{blue}, corresponding to two possible values of some physical observable. The precise interpretation of this observable will not matter, and we will rely only on the fact of it taking two distinct values. 

Observers may exist in these domains with some probability, where we assume this probability is independent of the color of any domain. As outlined in section~\ref{SEC:Generalities}, we characterize this probability by first introducing a vector of observer occupancy (or a `configuration') via the notation $\vec{\sigma} := (\sigma_{1}, \sigma_{2}, \dots, \sigma_{N})$, where, for $i=1,2,\dots,N$, $\sigma_{i}=1$ denotes the existence of observers in domain $i$, and $\sigma_{i}=0$ denotes there are no observers in domain $i$. There will, of course, be $2^N$ such configurations $\vec{\sigma}$, the set of which we label $\mathcal{K}$.

We will further assume that the probability of observers in a domain is independent of that for all the other domains, but also that these probabilities are \emph{not} generally the same.\footnote{The general results below (in section~\ref{SEC:Results}) do not depend on any assumption that the different domains receive independent but unequal (or even equal) probabilities. However, it is natural to assume that $(a)$ the existence of observers in different domains constitute probabilistically independent events (e.g., due to separate processes of evolution), and $(b)$ that these probabilities can be unequal, reflecting the fact that different domains can vary in their hostility to life.} This implies that the probability of $\vec{\sigma}$, denoted by $P(\vec{\sigma})$, factorizes into a product of marginals $P_{i}(\sigma_{i})$:  $P(\vec{\sigma}) = \prod_{i=1}^{N}P_{i}(\sigma_{i})$. If we let $p_{i}$ denote the probability of the existence of observers in domain $i$, then $(1-p_{i})$ is the probability of no observers in that domain, and since $\sigma_{i} = \textrm{1 or 0}$, we can write $P_{i}(\sigma_{i}) = p_{i}^{\sigma_{i}}(1-p_{i})^{1-\sigma_{i}}$, giving
\begin{equation}\label{EQN:Prob_Joint}
P(\vec{\sigma}) = \prod_{i=1}^{N}p_{i}^{\sigma_{i}}(1-p_{i})^{1-\sigma_{i}}.
\end{equation}

As to our data, we assume: we exist within one of these domains and observe the color red. That is, our data $D_{0}$ is
\begin{itemize}
\item[$D_{0}$:] There exists a domain with observers in it who see \emph{red}.
\end{itemize}

To write down the first-person likelihood for our data in accord with Eq.~(\ref{EQN:1stPerson}), we also need to specify the theories and the xerographic distributions we are interested in. 

The theories we consider are ones that specify the color of each of the $N$ domains. In particular, each theory will be denoted by a vector $\mathcal{T} = (T_{1}, T_{2}, \dots, T_{N})$, where, for $i=1,2,\dots,N$,  $T_{i} = 1$  when the theory predicts that domain $i$ is red, and $T_{i} = 0$ when the theory predicts it is blue. 

The xerographic distributions we consider will be defined to take the value zero, for each domain outside some nonempty subset (among all possible (nonempty) subsets) of domains $\mathcal{V}=\{1,2,\dots,N\}$ in the multiverse. To fix notation, if we let $\mathcal{C}$ denote the power set of $\mathcal{V}$ excluding the empty set, i.e., $\mathcal{C} = 2^{\mathcal{V}}\backslash \emptyset$, then a nonempty subset of domains $c\in\mathcal{C}$, outside which a xerographic distribution must be zero, will be denoted by $c=\{v_{1}, v_{2}, \dots, v_{M}\}$ (where, of course, $1\leq M\leq N$). The $v_{\alpha}$'s are simply integers labeling different domains in $\mathcal{V}$. We will explicitly write xerographic distributions as $\xi_{c}$, with the subscript $c$ indicating the subset over which they can be nonzero (so this subset $c$ is a superset of the support of a xerographic distribution). For the sake of simplicity, we will assume that the elements of $c$ are listed in increasing order, though nothing physical in what follows will depend on this. 

Then for each possible $c$, our xerographic distributions will fall into two classes:
\begin{itemize}
\item[{\bf C1}] The first class of xerographic distributions assumes that we are typical among instances of observers who see our data $D_{0}$, and will be denoted by $\xi^{\textrm{typD}}_c$. 
\item[{\bf C2}] The second class of xerographic distributions assumes we are typical among  instances of observers, regardless of what data they see (i.e., regardless of whether they measure red or blue for their particular domain). These distributions will be denoted by $\xi^{\textrm{typO}}_c$.
\end{itemize}

Note that only \emph{one} of the xerographic distributions imposes {\bf PM}, namely, $\xi^{\textrm{PM}}:=\xi^{\textrm{typD}}_\mathcal{V}$. That is, the principle of mediocrity is represented by the distribution that assumes we are typical among \emph{all} instances of our data over the multiverse (note that $c = \mathcal{V}$ for this distribution). Under the assumptions laid out in section~\ref{SEC:Testing_typicality}, any other xerographic distribution encodes some form of \emph{non-mediocrity}.

With these preliminaries in mind, we can construct the analog of Eq.~(\ref{EQN:1stPerson}) for our multiverse model. We will do so separately for {\bf C1} and {\bf C2} above. We turn first to {\bf C2}: where we will construct the first-person likelihood for our data $D_{0}$, assuming we are typical among instances of observers, regardless of their data, as encoded by $\xi^{\textrm{typO}}_c$.

\subsection{Construction of $P^{(1p)}(D_{0}|\mathcal{T}, \xi^{\textrm{typO}}_c)$}\label{SEC:Construction_TypO}

As indicated in the outer sum of Eq.~(\ref{EQN:1stPerson}), the definition of the first-person likelihood for our data $D_{0}$, sums over the subset $\mathcal{K}_{D_{0}}(\mathcal{T})$ of configurations that contain our data, according to theory $\mathcal{T}$. Keeping in mind that any xerographic distribution must be zero outside a nonempty subset $c=\{v_{1}, v_{2}, \dots, v_{M}\}$ of multiverse domains $\mathcal{V}$, we let $\mathcal{K}_{D_{0}}(c,\mathcal{T})\subset\mathcal{K}$ denote the corresponding subset, i.e. the subset of all configurations $\vec{\sigma}$ that contain instances of our data $D_{0}$, according to theory $\mathcal{T}$, within the domains specified by $c$. 

Next we need to characterize the xerographic distribution $\xi^{\textrm{typO}}_c$. For a configuration $\vec{\sigma}\in\mathcal{K}_{D_{0}}(c,\mathcal{T})$, the total number of observers within $c=\{v_{1}, v_{2}, \dots, v_{M}\}$ is simply $\sum_{\alpha = 1}^{M}\sigma_{v_{\alpha}}$; and the xerographic distribution corresponding to $\xi^{\textrm{typO}}_c$, assigns to the location $x_{l}$ of each observer-system in $c$, the value $\xi_{l} = 1/(\sum_{\alpha = 1}^{M}\sigma_{v_{\alpha}})$ (recalling that a location in our model is simply a multiverse domain). In the computation of the first-person likelihood, we sum the xerographic distribution over only those locations that contain instances of our data $D_{0}$. So to that end, following our notation in Eq.~(\ref{EQN:1stPerson}), for each configuration $\vec{\sigma}\in\mathcal{K}_{D_{0}}(c,\mathcal{T})$, let $L_{D_{0}}(c, \vec{\sigma}, \mathcal{T})$ denote the subset of locations in $c$,  where our data $D_{0}$ exists, according to $\mathcal{T}$. Note that the total number of instances of our data in $c$ is the just the size of this set $|L_{D_{0}}(c, \vec{\sigma}, \mathcal{T})|$. This can be written as $|L_{D_{0}}(c, \vec{\sigma}, \mathcal{T})| = \sum_{\beta = 1}^{M}\sigma_{v_{\beta}}T_{v_{\beta}}$: the dummy variable $v_{\beta}$ on the right hand side shows the $c$-dependence; and although there is no explicit $D_{0}$ dependence, recall that $T_{v_{\beta}}=1$ iff domain $v_{\beta}$ is red, which, when $\sigma_{v_{\beta}}=1$, corresponds to the existence of our data $D_{0}$ in domain $v_{\beta}$. 

We can now calculate a closed-form expression for $P^{(1p)}(D_{0}|\mathcal{T}, \xi^{\textrm{typO}}_c)$, by directly adapting Eq.~(\ref{EQN:1stPerson}) for the particular case considered here:
\begin{widetext}
\begin{eqnarray}
P^{(1p)}(D_{0}|\mathcal{T}, \xi^{\textrm{typO}}_c) &=& \sum_{\vec{\sigma}\in\mathcal{K}_{D_{0}}(c,\mathcal{T})} P(\vec{\sigma}) \sum_{l'\in L_{D_{0}}(c, \vec{\sigma},\mathcal{T})}\xi_{l'} \nonumber \\
&=& \sum_{\vec{\sigma}\in\mathcal{K}_{D_{0}}(c,\mathcal{T})} \prod_{i=1}^{N}p_{i}^{\sigma_{i}}(1-p_{i})^{1-\sigma_{i}} \sum_{l'\in L_{D_{0}}(c, \vec{\sigma},\mathcal{T})}\xi_{l'} \nonumber \\
&=& \sum_{\vec{\sigma}\in\mathcal{K}_{D_{0}}(c,\mathcal{T})} \prod_{i=1}^{N}p_{i}^{\sigma_{i}}(1-p_{i})^{1-\sigma_{i}} \sum_{l'\in L_{D_{0}}(c, \vec{\sigma},\mathcal{T})}\frac{1}{\sum_{\alpha = 1}^{M}\sigma_{v_{\alpha}}} \nonumber \\
&=& \sum_{\vec{\sigma}\in\mathcal{K}_{D_{0}}(c,\mathcal{T})} \prod_{i=1}^{N}p_{i}^{\sigma_{i}}(1-p_{i})^{1-\sigma_{i}} \frac{1}{\sum_{\alpha = 1}^{M}\sigma_{v_{\alpha}}} |L_{D_{0}}(c, \vec{\sigma}, \mathcal{T})| \nonumber \\
&=& \sum_{\vec{\sigma}\in\mathcal{K}_{D_{0}}(c,\mathcal{T})} \prod_{i=1}^{N}p_{i}^{\sigma_{i}}(1-p_{i})^{1-\sigma_{i}} \left(\frac{\sum_{\beta = 1}^{M} \sigma_{v_{\beta}}T_{v_{\beta}}}{\sum_{\alpha = 1}^{M} \sigma_{v_{\alpha}}}\right); \label{EQN:1stPersonO}
\end{eqnarray}
\end{widetext}
where we have used Eq.~(\ref{EQN:Prob_Joint}) to obtain the second line, and the fact that the value of the xerographic distribution is the same at each location ${l'\in L_{D_{0}}(c, \vec{\sigma},\mathcal{T})}$ to obtain the fourth line. 

\subsection{Construction of $P^{(1p)}(D_{0}|\mathcal{T}, \xi^{\textrm{typD}}_c)$}\label{SEC:Construction_TypD}

We turn next to {\bf C1}, and the construction of the first-person likelihood for our data $D_{0}$, assuming we are typical among instances of our data, as encoded by $\xi^{\textrm{typD}}_c$.
Having completed the above construction, it is straightforward to compute this. For a configuration $\vec{\sigma}\in\mathcal{K}_{D_{0}}(c,\mathcal{T})$, since the total number of observers who see our data within $c$ is $\sum_{\alpha = 1}^{M}\sigma_{v_{\alpha}}T_{v_{\alpha}}$, the xerographic distribution corresponding to $\xi^{\textrm{typD}}_c$ is $\xi_{l} = 1/(\sum_{\alpha = 1}^{M}\sigma_{v_{\alpha}}T_{v_{\alpha}})$ for each location $x_{l}$ of our data in $c$. The term that imposes the xerographic distribution in the computation of the first-person likelihood, namely $\sum_{l'\in L_{D_{0}}(c, \vec{\sigma},\mathcal{T})}\xi_{l'}$, is then simply unity: $\sum_{l'\in L_{D_{0}}(c, \vec{\sigma},\mathcal{T})}\xi_{l'}=1$. Hence, we find
\begin{equation}\label{EQN:1stPersonD}
P^{(1p)}(D_{0}|\mathcal{T}, \xi^{\textrm{typD}}_c) =  \sum_{\vec{\sigma}\in\mathcal{K}_{D_{0}}(c,\mathcal{T})} \prod_{i=1}^{N}p_{i}^{\sigma_{i}}(1-p_{i})^{1-\sigma_{i}}.  
\end{equation}
Note that we recover the analog of the claim in the last paragraph of section~\ref{SEC:Generalities}: that for {\bf C1}, the first-person likelihood $P^{(1p)}(D_{0}|\mathcal{T}, \xi^{\textrm{typD}}_c)$, is equal to the appropriate third-person likelihood---which expresses the likelihood that there is at least one observing system that witnesses our data $D_{0}$, among the multiverse domains specified by $c$.

Equations~(\ref{EQN:1stPersonO}) and~(\ref{EQN:1stPersonD}) are the appropriate generalizations of the likelihoods calculated for the cyclic cosmological model of~\citet{srednicki+hartle_10}. In order to make this connection more precise, and to set the stage for some of the computations described in section~\ref{SEC:Results}, we now show that in the case where our multiverse model reduces to the cosmological model of~\citet{srednicki+hartle_10}, Eqs.~(\ref{EQN:1stPersonO}) and~(\ref{EQN:1stPersonD}) indeed reduce to their likelihoods. 
 
\subsection{A multiverse of cycles}\label{SEC:SH_multiverse}

Consider, then, the case where the $N$ domains of the multiverse model introduced in section~\ref{SEC:Model_preliminaries} are stretched out in time, so that the index $i$ that labels a domain, corresponds to the order in which the domain occurs in what we will refer to as a `multiverse of cycles'. Assume further that the xerographic distribution can be nonzero for each of these domains, so that $c=\mathcal{V}$. Furthermore, assume, as in~\cite{srednicki+hartle_10}, that the probability of observers existing in each of these cycles is the same (as well as independent of whether they exist in other cycles), and is given by $p_{i} = p$. We wish to first calculate the likelihood of our data under the assumption that we are typical among all instances of observers, i.e., $P^{(1p)}(D_{0}|\mathcal{T}, \xi^{\textrm{typO}}_\mathcal{V})$. Under these assumptions, Eq.~(\ref{EQN:1stPersonO}) reduces to
\begin{widetext}
\begin{equation}
P^{(1p)}(D_{0}|\mathcal{T}, \xi^{\textrm{typO}}_\mathcal{V}) =
  \sum_{\vec{\sigma}\in\mathcal{K}_{D_{0}}(\mathcal{V},\mathcal{T})} p^{\sum_{i=1}^{N}\sigma_{i}}(1-p)^{N-\sum_{j=1}^N\sigma_{j}}\left(\frac{\sum_{m = 1}^{N} \sigma_{m}T_{m}}{\sum_{n = 1}^{N} \sigma_{n}}\right).
\label{EQN:1stPersonOHS}
\end{equation}
\end{widetext}

The large amount of symmetry in this expression allows us to organize the outermost sum by separately considering configurations $\vec{\sigma}$ according to the total number of observers $\sum_{i=1}^{N}\sigma_{i}$ in each configuration. We will call the total number of observers in each configuration $n_{O}$, and note that this will range from 1 to $N$ for consistency with the fact that there exists at least one cycle with observers, that is, for consistency with our data (recall that we are not counting how many individuals are in each domain, just whether domains house observers or not). Equation~(\ref{EQN:1stPersonOHS}) can then be thought of as $P^{(1p)}(D_{0}|\mathcal{T}, \xi^{\textrm{typO}}_\mathcal{V}) = \sum_{n_{O}=1}^{N}F(p,n_{O}, N_{R})$, for some function $F$ we will compute below, where $N_{R}$ denotes the total number of red cycles predicted by the theory $\mathcal{T}$. There are two obvious cases to consider in this sum: namely, $1 \leq n_{O} \leq  N_{R}$ and $N_{R} < n_{O} \leq  N$, and we formally treat these two cases separately. 

For $1 \leq n_{O} \leq  N_{R}$, one can generate an expression for $F(p, n_{O}, N_{R})$ by sequentially considering all possible numbers of observers (out of a maximum of $n_{O}$ observers) placed in $N_{R}$ red cycles. In general, one finds
\begin{widetext}
\begin{eqnarray}
F(p,n_{O},N_{R})&=& p^{n_{O}}(1-p)^{N-n_{O}}\frac{1}{n_{O}}\left[\binom{N_{R}}{n_{O}} n_{O} +
 \binom{N_{R}}{n_{O}-1}\binom{N-N_{R}}{1} (n_{O}-1)+\cdots+\binom{N_{R}}{1}\binom{N-N_{R}}{n_{O}-1}\right]\nonumber \\
&=& p^{n_{O}}(1-p)^{N-n_{O}}\frac{1}{n_{O}}\sum_{k=0}^{n_{O}-1}\binom{N_{R}}{n_{O}-k}\binom{N-N_{R}}{k}(n_{O}-k)\nonumber \\
&=& p^{n_{O}}(1-p)^{N-n_{O}}\frac{1}{n_{O}}N_{R}\sum_{k=0}^{n_{O}-1}\binom{N-N_{R}}{k}\binom{N_{R}-1}{n_{O}-1-k}\nonumber \\
&=& p^{n_{O}}(1-p)^{N-n_{O}}\frac{1}{n_{O}}N_{R}\binom{N-1}{n_{O}-1}\nonumber \\
&=& p^{n_{O}}(1-p)^{N-n_{O}}N_{R}\frac{1}{N}\binom{N}{n_{O}}, 
\end{eqnarray}
\end{widetext}
where we have used a standard binomial identity to obtain the third and fifth lines, together with Vandermonde's identity to obtain the fourth one. 

In the case where $N_{R} < n_{O} \leq  N$, it is not difficult to show that we obtain precisely the same final result using a similar sequence of steps. Putting this all together, gives
\begin{eqnarray}
P^{(1p)}(D_{0}|\mathcal{T}, \xi^{\textrm{typO}}_\mathcal{V}) &=&\frac{N_{R}}{N}\sum_{n_{O}=1}^{N}\binom{N}{n_{O}}p^{n_{O}}(1-p)^{N-n_{O}}\nonumber \\
&=& \frac{N_{R}}{N}\left[1-(1-p)^{N}\right],\label{EQN:FP_SH_Nr_typO}
\end{eqnarray}
agreeing with the corresponding expression (Eqs. (5.8) and (B5)) in~\cite{srednicki+hartle_10}. 

A similar calculation (presented in the appendix) gives the correct formula for $P^{(1p)}(D_{0}|\mathcal{T}, \xi^{\textrm{PM}})$ (recall that $\xi^{\textrm{PM}}:=\xi^{\textrm{typD}}_\mathcal{V}$ ), where
\begin{equation}\label{EQN:FP_SH_Nr}
P^{(1p)}(D_{0}|\mathcal{T}, \xi^{\textrm{PM}}) = 1-(1-p)^{N_{R}},
\end{equation}
corresponding to Eq. (5.5) in~\cite{srednicki+hartle_10}. In this way, the generalized multiverse model introduced in section~\ref{SEC:Model_preliminaries} reduces to the cyclic model of~\citet{srednicki+hartle_10} under the appropriate simplifying assumptions.

\section{Evaluating the principle of mediocrity}\label{SEC:Results}

With the results of the last section in hand, we can now address the central conceptual task of this paper: namely, to assess the predictive power (understood in Bayesian terms) of the principle of mediocrity, under the scheme introduced in section~\ref{SEC:Xerographic_Distributions}. We will develop the answers to three central questions in turn: (A) What framework  produces the highest likelihoods? (B) For a fixed theory, what xerographic distribution gives rise to the framework with the highest likelihoods? (C) Does the principle of mediocrity generally provide the framework with the highest likelihoods?

\subsection{The best performing framework}\label{SEC:Best_Performing}

To begin our analysis, we show that the framework whose likelihood attains the supremum of the likelihoods for all frameworks considered, and for all assignments of probabilities $\{p_{i}\}_{i=1}^{N}$ to multiverse domains, is $(\mathcal{T}_{\textrm{all red}}, \xi^{\textrm{PM}})$: that is, the theory that predicts each of the domains is red ($\mathcal{T}=\mathcal{T}_{\textrm{all red}}$), together with the xerographic distribution corresponding to the principle of mediocrity ($\xi=\xi^{\textrm{PM}}:=\xi^{\textrm{typD}}_\mathcal{V}$). For this framework, the likelihood attained is the same as that with a xerographic distribution corresponding to typicality across all observers for $c=\mathcal{V}$, that is, for $\xi=\xi^{\textrm{typO}}_\mathcal{V}$. 

To show this, let the $p_{i}$'s be arbitrarily chosen but fixed. Note first that for arbitrary $c$ and $\mathcal{T}$, Eqs.~(\ref{EQN:1stPersonO}) and~(\ref{EQN:1stPersonD}) imply $P^{(1p)}(D_{0}|\mathcal{T}, \xi^{\textrm{typO}}_{c}) \leq P^{(1p)}(D_{0}|\mathcal{T}, \xi^{\textrm{typD}}_{c})$: as the same configurations $\vec{\sigma}\in\mathcal{K}_{D_{0}}(c,\mathcal{T})$ contribute to the sums in Eqs.~(\ref{EQN:1stPersonO}) and~(\ref{EQN:1stPersonD}), but in the case of Eq.~(\ref{EQN:1stPersonO}), each contributing term is multiplied by a factor that is less than or equal to 1. Now, the \emph{maximal} value of $P^{(1p)}(D_{0}|\mathcal{T}, \xi^{\textrm{typD}}_{c})$ is attained when we choose $c$ and $\mathcal{T}$ such that $\mathcal{K}_{D_{0}}(c,\mathcal{T})$ includes the maximal number of configurations in the sum over (manifestly non-negative) probabilities of configurations. This will occur for the case where all multiverse domains are included ($c=\mathcal{V}$) and for the theory that predicts that all domains are red. Note finally that for $\mathcal{T} =\mathcal{T}_{\textrm{all red}}\equiv(1,1,\dots,1)$, $\sum_{\beta = 1}^{M} \sigma_{v_{\beta}}T_{v_{\beta}} = \sum_{\alpha = 1}^{M} \sigma_{v_{\alpha}}$ (regardless of $c$), and so the final term in brackets in Eq.~(\ref{EQN:1stPersonO}) is unity;  hence the likelihood for the framework $(\mathcal{T}_{\textrm{all red}}, \xi^{\textrm{PM}})$, coincides with the likelihood for $(\mathcal{T}_{\textrm{all red}}, \xi^{\textrm{typO}}_\mathcal{V})$. 

\begin{figure*}
\includegraphics[width=0.45\linewidth]{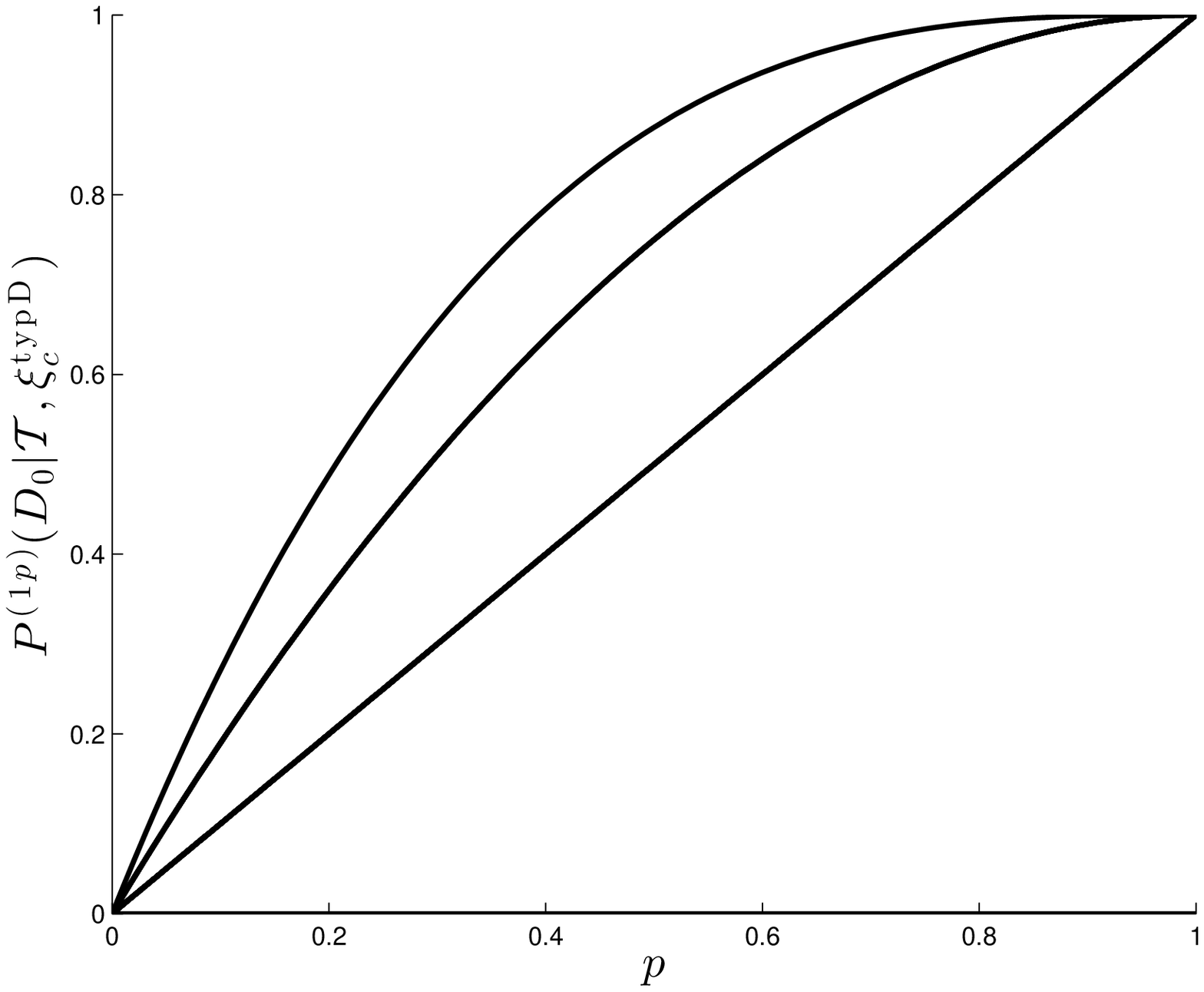}
\hfill
\includegraphics[width=0.45\linewidth]{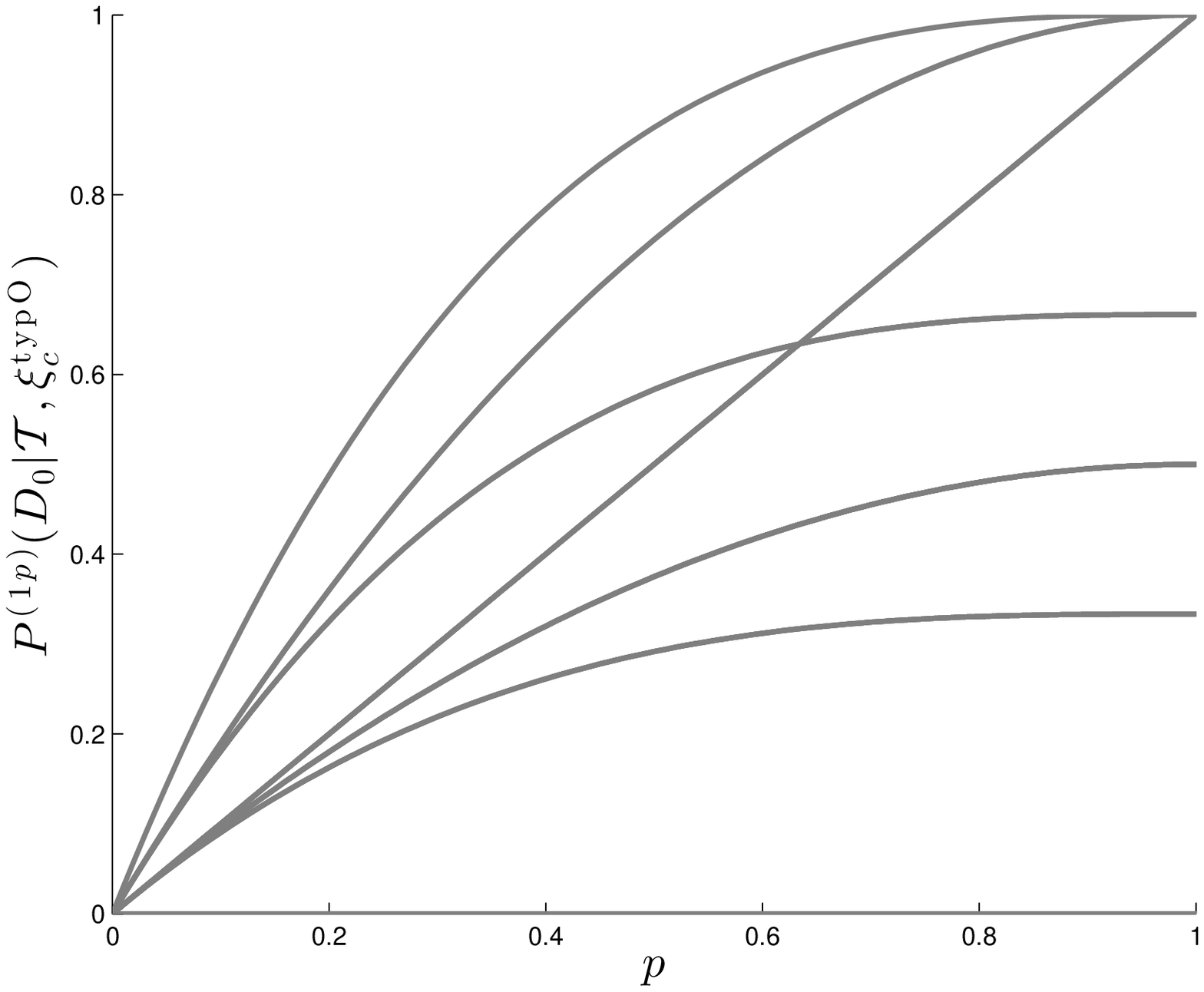}
\caption{(Left): First-person likelihoods for our data $D_{0}$, given a theory $\mathcal{T}$ and a xerographic distribution $\xi^{\textrm{typD}}_{c}$, for a multiverse of cycles with a total of 3 cycles (see section~\ref{SEC:SH_multiverse}). The theories considered include all possible theories assigning the observable red/blue to each cycle (for a total of $2^{3}$ theories), and the xerographic distributions can be nonzero only up to (and including) some terminal cycle (corresponding to the final element of $c$). We implement the assumption that we are typical with respect to other instances of our data up to (and including) that terminal cycle (for a total of 3 possible xerographic distributions; see {\bf C1} of section~\ref{SEC:Model_preliminaries}). Likelihoods are shown as a function of $p$, the probability of the existence of observers in each cycle, which is here assumed to be the same for each cycle. Likelihoods for all possible frameworks (found by varying $\mathcal{T}$ and $c$---a total of $24$ frameworks) have been superimposed on the plot. In this case, the existence of equivalent functional forms for likelihoods for different frameworks, gives rise to 4 distinct traces (including likelihoods which are zero for all values of $p$). The uppermost trace corresponds \emph{only} to the framework $(\mathcal{T}_{\textrm{all red}}, \xi^{\textrm{PM}}:=\xi^{\textrm{typD}}_\mathcal{V})$, and is the supremum of the likelihoods for all possible values of $p$. (Right): This plot displays similar information as in the left panel, but for the xerographic distribution that assumes we are typical of all observers, regardless of what data they see ({\bf C2} of section~\ref{SEC:Model_preliminaries}). Here, the existence of equivalent functional forms for likelihoods for different frameworks, gives rise to 7 distinct traces (again, including likelihoods which are zero for all values of $p$). The uppermost trace is the same as the uppermost trace for the left panel, and corresponds only to the framework  $(\mathcal{T}_{\textrm{all red}}, \xi^{\textrm{typO}}_\mathcal{V})$.}
\label{FIG:SH_N3_likelihoods}
\end{figure*}

To see how these conclusions manifest in a particular setting, consider the multiverse of cycles introduced in section~\ref{SEC:SH_multiverse}, which assumes that the probability of observers existing in any cycle is the same for each cycle and is given by $p$. We will consider the case where the xerographic distribution can be nonzero only on an initial segment of cycles starting with the first cycle and ending with some terminal cycle (after which the xerographic distribution is assumed to be zero). So for any number of cycles $N$, there are a total of $N$ such possible `cut-off schemes'. Figure~\ref{FIG:SH_N3_likelihoods} shows likelihoods as a function of $p$ for the case where $N=3$.
We see that the highest likelihoods indeed occur for the framework with $(\mathcal{T}_{\textrm{all red}}, \xi^{\textrm{PM}})$ (uppermost trace in Fig.~\ref{FIG:SH_N3_likelihoods} (left)), and that these likelihoods coincide with the likelihoods for $(\mathcal{T}_{\textrm{all red}}, \xi^{\textrm{typO}}_\mathcal{V})$ (uppermost trace in Fig.~\ref{FIG:SH_N3_likelihoods} (right)). 

\subsection{The best xerographic distribution for a fixed theory}

It is an encouraging check of our intuition that the framework with the highest likelihood (and so also the highest posterior probability under the assumptions of section~\ref{SEC:Bayesian}) involves the theory that predicts all domains are red. A natural question that arises is: for a \emph{fixed theory}, what xerographic distribution leads to the best performing framework? It does not take much more work to show that under these circumstances, the framework whose likelihood attains the supremum of the likelihoods over all xerographic distributions considered, for all assignments of probabilities $p_{i}$ to the multiverse domains, is the one whose xerographic distribution corresponds to {\bf PM}: $\xi=\xi^{\textrm{PM}}:=\xi^{\textrm{typD}}_\mathcal{V}$. 

To prove this, fix the theory $\mathcal{T}$ and let the $p_{i}$'s be arbitrarily chosen but fixed. For an arbitrary $c$, we showed in section~\ref{SEC:Best_Performing} that $P^{(1p)}(D_{0}|\mathcal{T}, \xi^{\textrm{typO}}_{c}) \leq P^{(1p)}(D_{0}|\mathcal{T}, \xi^{\textrm{typD}}_{c})$. Now, as in the proof there, the choice of $c$ that will maximize $P^{(1p)}(D_{0}|\mathcal{T}, \xi^{\textrm{typD}}_{c})$ is the one that will include the most number of non-negative terms in the sum in Eq.~(\ref{EQN:1stPersonD}). This is just $c=\mathcal{V}$, and hence ($\mathcal{T}, \xi^{\textrm{PM}}$) will in general give the highest likelihood. 

We see how this claim manifests in a simple case, by again considering the multiverse of cycles introduced in section~\ref{SEC:SH_multiverse}, with xerographic distributions implementing a cut-off in time, as described in the last paragraph of section~\ref{SEC:Best_Performing}.
\begin{figure*}
\includegraphics[width=0.95\linewidth]{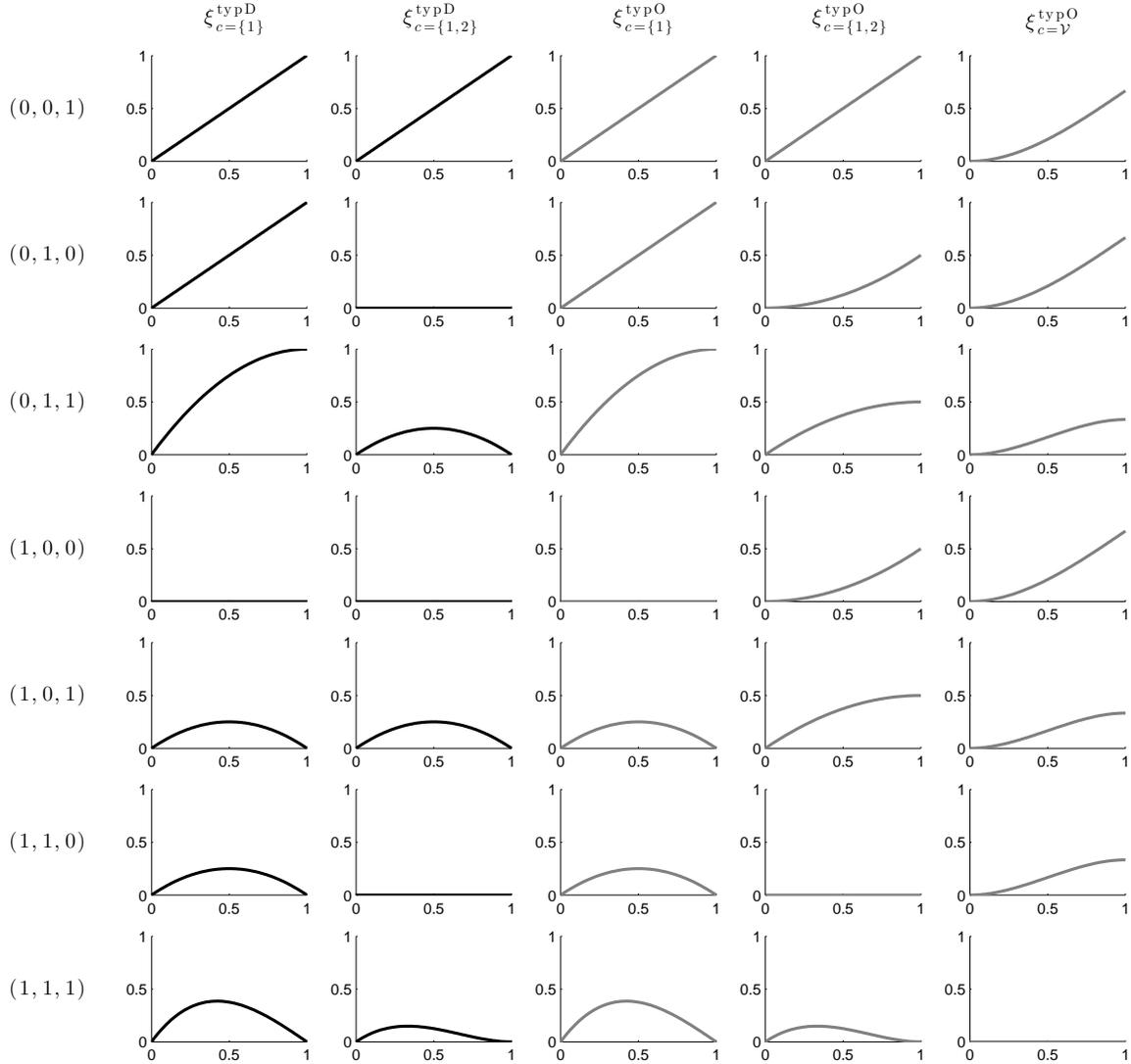}
\hfill
\caption{The multiverse of cycles introduced in section~\ref{SEC:SH_multiverse} with $N=3$ cycles. Each subplot shows $[P^{(1p)}(D_{0}|\mathcal{T}, \xi^{\textrm{PM}}) - P^{(1p)}(D_{0}|\mathcal{T}, \xi^{\textrm{typX}}_{c})]$ vs $p$ (where $\textrm{X}$ is either $\textrm{D}$ or $\textrm{O}$). The term in square brackets is the difference between the first-person likelihood for the framework that implements the principle of mediocrity, namely ($\mathcal{T}, \xi^{\textrm{PM}}:=\xi^{\textrm{typD}}_{\mathcal{V}}$), and the likelihood for another framework specified by varying only the xerographic distribution in accord with the cut-off procedure described in the last paragraph of section~\ref{SEC:Best_Performing}. The variable $p$ (labeling the $x$-axis of each subplot) is the probability of the occurrence of observers in any cycle (and is assumed to be the same in each cycle). The theory is shown on the left of each row, with 1 or 0 respectively denoting the prediction of red or blue in the corresponding cycle. The xerographic distribution against which we are comparing the principle of mediocrity is shown on the top of each column (traces referring to $\xi^{\textrm{typO}}_{c}$ are shown in grey to guide the eye). We see that for each row, that is, for a fixed theory, the difference is non-negative for all possible values of $p$. It is in this sense that we claim the best xerographic distribution for a fixed theory is that corresponding to the principle of mediocrity.
}
\label{FIG:SH_N3_FIxTheory}
\end{figure*}
Figure~\ref{FIG:SH_N3_FIxTheory} displays plots of the difference between the likelihood for the framework with some theory $\mathcal{T}$ and xerographic distribution corresponding to the principle of mediocrity $\xi=\xi^{\textrm{PM}}:=\xi^{\textrm{typD}}_{\mathcal{V}}$, and the likelihood for the framework with the same theory and all other xerographic distributions (the probability $p$ of observers in domains, labels the $x$-axis of each subplot). The case for $N=3$ cycles is displayed. We see that for each row of the plot (corresponding to a fixed theory), the difference is non-negative in each case, confirming the general claim advanced in this section. 

\subsection{Does the principle of mediocrity generally give rise to the most predictive frameworks?}\label{SEC:PM_WBS}

In light of the results in the last two subsections it is natural to ask whether there are cases where {\bf PM} does not provide the highest likelihoods. We will show by construction that indeed it does \emph{not}, when one is allowed to vary both the underlying theory and the xerographic distribution. For the sake of simplicity, we will restrict attention to the multiverse of cycles introduced in section~\ref{SEC:SH_multiverse}, where again, the probability of observers in any cycle is the same for each cycle and is given by $p$. 

There is a strong motivation for embarking on this search, since some well-motivated theories suggest xerographic distributions which do not express the principle of mediocrity. We will further explicate such scenarios in the next section; but for now, we aim to figure out whether in such instances, likelihoods are generally highest for those frameworks that involve the principle of mediocrity.

To be explicit about the terms for our search: we are interested in whether for a given framework characterized by {\bf PM}, that is, for some `reference framework' ($\mathcal{T}, \xi^{\textrm{PM}}:=\xi^{\textrm{typD}}_\mathcal{V}$), there could exist another framework ($\mathcal{T}^{\star}, \xi^{\star}$) with a higher likelihood for at least some values of $p$. To steer the discussion away from more trivial cases, we will only consider situations where:
\begin{itemize}
\item[(i)] $\mathcal{T} \neq \mathcal{T}_{\textrm{all red}}$: since otherwise, by the results of section~\ref{SEC:Best_Performing}, we will not be able to find the required framework ($\mathcal{T}^{\star}, \xi^{\star}$).
\item[(ii)] $\mathcal{T} \neq \mathcal{T}_{\textrm{no red}}$: where $\mathcal{T}_{\textrm{no red}} = (0,0,\dots,0)$ is the theory that predicts no red cycles; since then the likelihood of our data $D_{0}$, given the framework ($\mathcal{T}_{\textrm{no red}}, \xi^{\textrm{PM}}$), vanishes, and we will be able to find the required frameworks ($\mathcal{T}^{\star}, \xi^{\star}$) rather trivially.
\item[(iii)] $\xi^{\star}\neq \xi^{\textrm{PM}}$: since we already know from section~\ref{SEC:Best_Performing} that for $\xi^{\star} = \xi^{\textrm{PM}}$, $\mathcal{T}^{\star} = \mathcal{T}_{\textrm{all red}}$ will provide the supremum of the likelihoods in this case---and moreover, we are interested in finding frameworks where {\bf PM} is not integral to constructing higher likelihoods.
\item[(iv)] $(\mathcal{T}^{\star}, \xi^{\star}) \neq (\mathcal{T}_{\textrm{all red}}, \xi^{\textrm{typO}}_\mathcal{V})$: since again, this corresponds to the supremum as mentioned in (iii), following the results of section~\ref{SEC:Best_Performing}.
 \end{itemize}

So consider the case $N=3$. Set $\mathcal{T}$ in our reference framework ($\mathcal{T}, \xi^{\textrm{PM}}$), to be the theory that predicts a total of one of the three cycles is red: $\mathcal{T}=\mathcal{T}_{\textrm{one\,red}}$. Then, from Eq.~(\ref{EQN:FP_SH_Nr}), we have $P^{(1p)}(D_{0}|\mathcal{T}_{\textrm{one red}}, \xi^{\textrm{PM}}) = p$. It is straightforward to show from Eq.~(\ref{EQN:1stPersonO}) that for the theory that predicts only the first two cycles are red, which we will denote by $\mathcal{T}^{\star}=(1,1,0)\equiv\mathcal{T}_{RRB}$, the xerographic distribution given by $\xi^{\textrm{typO}}_{c=\{1,2\}}$, which (clearly) \emph{does not} correspond to the principle of mediocrity, implies: $P^{(1p)}(D_{0}|\mathcal{T}_{RRB}, \xi^{\textrm{typO}}_{c=\{1,2\}}) = p(2-p)$. So the framework $(\mathcal{T}_{RRB}, \xi^{\textrm{typO}}_{c=\{1,2\}})$ has a  likelihood higher than the framework involving {\bf PM}, since $p(2-p) > p$ (for all non-trivial values of $p$, i.e. $p\in (0,1)$).

\begin{figure*}
\includegraphics[width=0.45\linewidth]{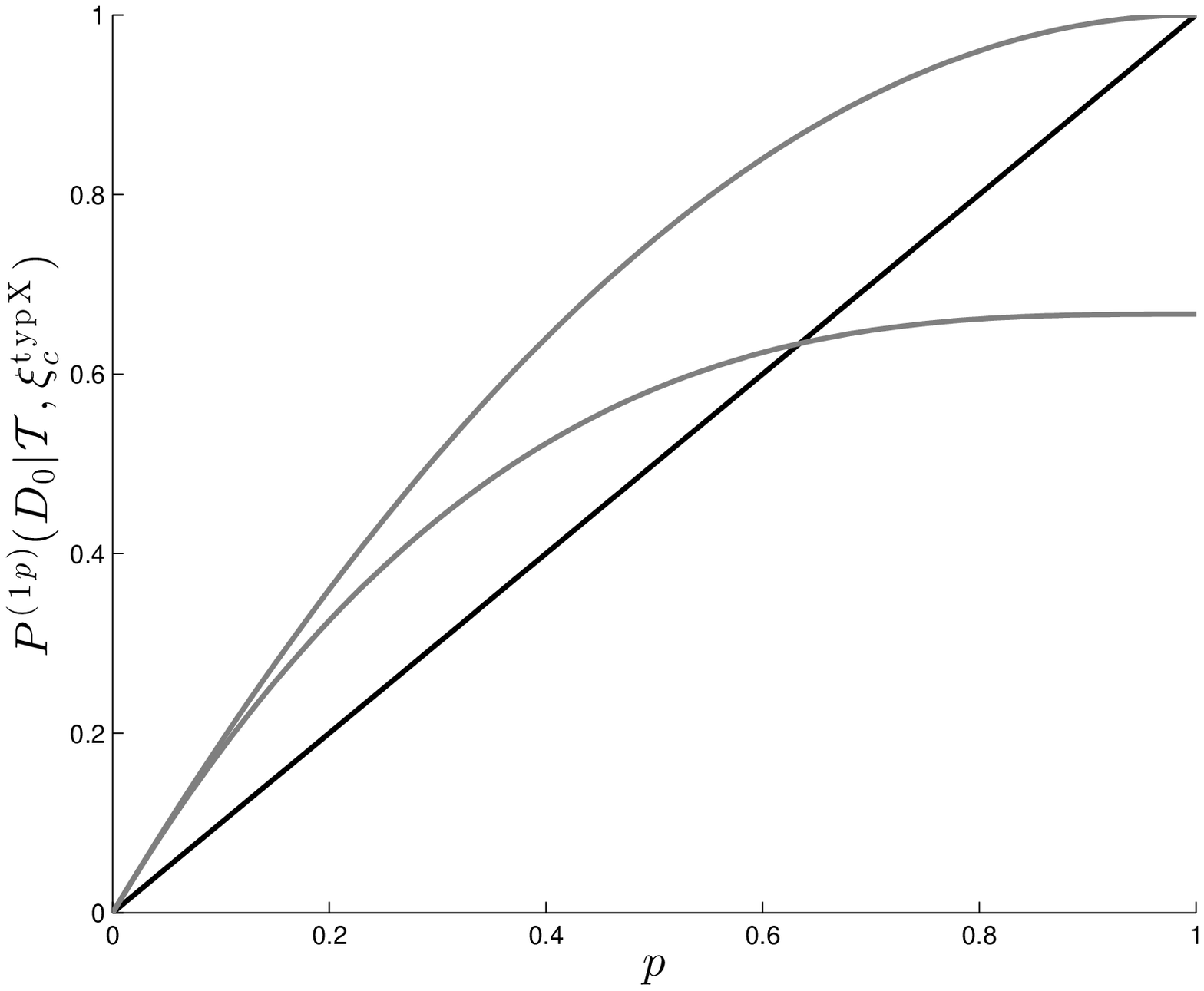}
\hfill
\includegraphics[width=0.45\linewidth]{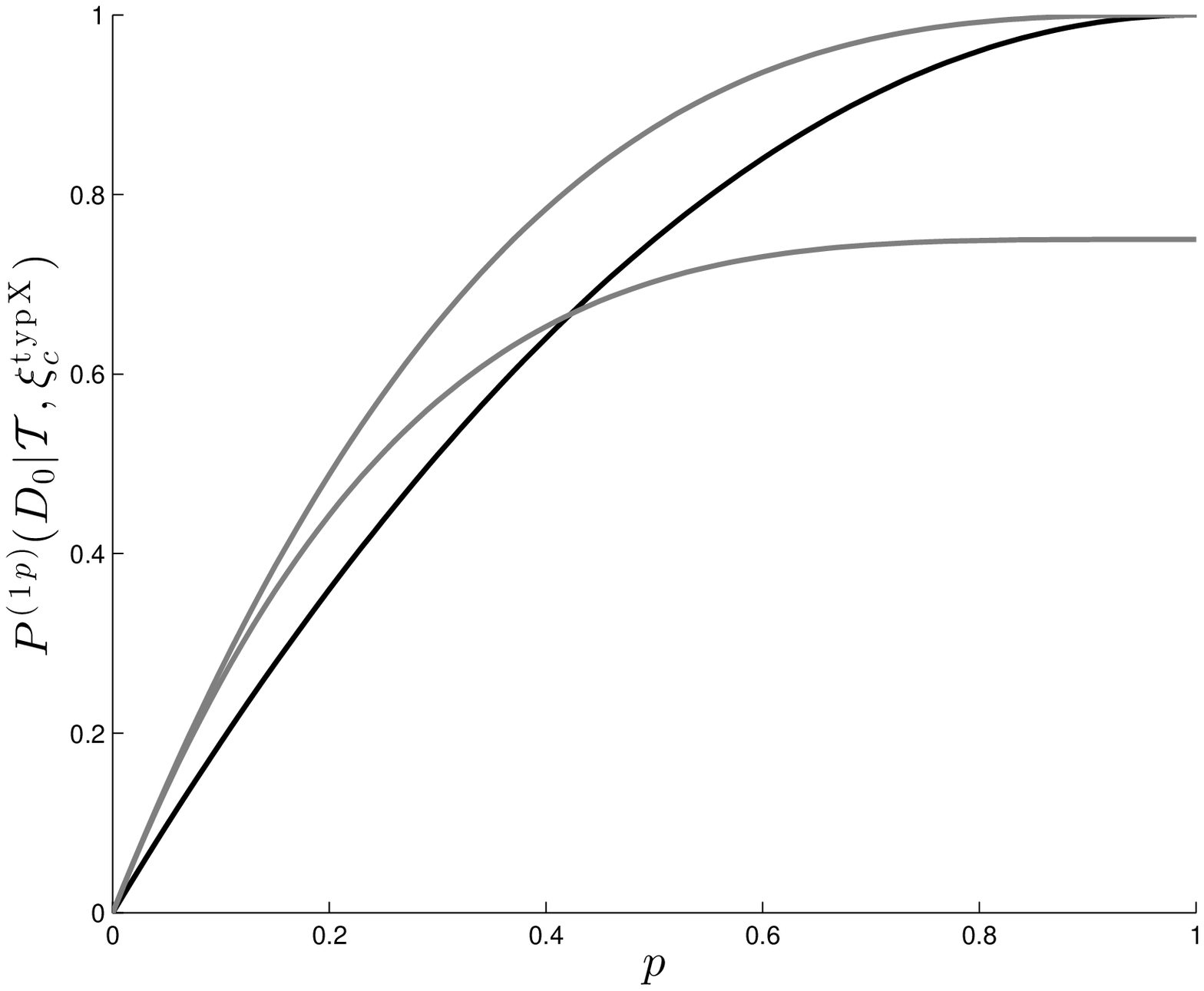}
\caption{(Left): First-person likelihoods for our data $D_{0}$, given a particular theory $\mathcal{T}$ and xerographic distributions $\xi^{\textrm{typX}}_{c}$ (where $\textrm{X}$ is either $\textrm{D}$ or $\textrm{O}$), vs $p$, for the  multiverse of cycles introduced in section~\ref{SEC:SH_multiverse} for $N=3$. The likelihoods for three separate frameworks are shown. The black line shows the likelihood for the framework ($\mathcal{T}_{\textrm{one\;red}}, \xi^{\textrm{PM}}:=\xi^{\textrm{typD}}_\mathcal{V}$), where the theory predicts a single red cycle and the xerographic distribution corresponds to the principle of mediocrity [$P^{(1p)}(D_{0}|\mathcal{T}_{\textrm{one red}}, \xi^{\textrm{PM}}) = p$]. The uppermost grey trace corresponds to the framework $(\mathcal{T}_{RRB}, \xi^{\textrm{typO}}_{c=\{1,2\}})$, where the theory predicts only the first two cycles are red ($\mathcal{T}_{RRB}\equiv(1,1,0)$), and the xerographic distribution corresponds to typicality with respect to observers considered among only the first two cycles [$P^{(1p)}(D_{0}|\mathcal{T}_{RRB}, \xi^{\textrm{typO}}_{c=\{1,2\}}) = p(2-p)$]. We note that this framework \emph{does better} than that containing a xerographic distribution implementing the principle of mediocrity. The lower grey trace corresponds to the likelihood for the framework ($\mathcal{T}_{\textrm{two\,red}}, \xi^{\textrm{typO}}_{\mathcal{V}}$), whose theory predicts that some two of the cycles are red, and that we are typical with respect to observers considered over the entire multiverse [$P^{(1p)}(D_{0}|\mathcal{T}_{\textrm{two\,red}}, \xi^{\textrm{typO}}_{\mathcal{V}}) = \frac{2}{3}p(p^2-3p+3)$]. This trace crosses the likelihood for ($\mathcal{T}_{\textrm{one\;red}}, \xi^{\textrm{PM}}$), implying a parametric dependence of the dominance of the framework implementing the principle of mediocrity. The cross-over point of this dependence occurs for $p\sim 0.63$. (Right): Similar results for the case $N=4$. Here, the black trace shows the likelihood for the framework implementing the principle of mediocrity with $P^{(1p)}(D_{0}|\mathcal{T}_{\textrm{two\,red}}, \xi^{\textrm{PM}}) = p(2-p)$. The uppermost trace corresponds to $P^{(1p)}(D_{0}|\mathcal{T}_{\textrm{RRRB}}, \xi^{\textrm{typO}}_{c = \{1,2,3\}}) = p(p^{2}-3p+3)$ (where $\mathcal{T}_{RRRB}\equiv(1,1,1,0)$). The lower grey trace corresponds to the likelihood  $P^{(1p)}(D_{0}|\mathcal{T}_{\textrm{three\;red}}, \xi^{\textrm{typO}}_\mathcal{V}) = 3p(4-6p+4p^{2}-p^{3})/4$. In this latter case, the principle of mediocrity is only dominant for $p>\sim 0.42$.}
\label{FIG:Figure_3}
\end{figure*}

More interesting behavior can be exhibited for the case where we take $\mathcal{T}^{\star}$ to be the theory that predicts a total of two of the cycles are red and the xerographic distribution to be $\xi^{\textrm{typO}}_{\mathcal{V}}\equiv \xi^{\textrm{typO}}_{c=\{1,2,3\}}$. In this case, as seen from Eq.~(\ref{EQN:FP_SH_Nr_typO}),  $P^{(1p)}(D_{0}|\mathcal{T}_{\textrm{two\,red}}, \xi^{\textrm{typO}}_{\mathcal{V}}) = \frac{2}{3}p(p^2-3p+3)$. This likelihood displays the behavior that it does better than our reference framework ($\mathcal{T}_{\textrm{one\;red}}, \xi^{\textrm{PM}}$), depending on the value of $p$. In particular, for $p >\sim 0.63$, ($\mathcal{T}_{\textrm{one\,red}}, \xi^{\textrm{PM}}$) has a higher likelihood than ($\mathcal{T}_{\textrm{two\,red}}, \xi^{\textrm{typO}}_{\mathcal{V}}$), whereas the situation is reversed otherwise. Such behavior thereby exhibits parametric dependence of the success of {\bf PM}. Both this situation and the one discussed in the last paragraph are displayed graphically in Fig.~\ref{FIG:Figure_3} (left).

Similar results can be obtained in the case where $N=4$ (see Fig.~\ref{FIG:Figure_3} (right)). In this case, there exists a framework implementing {\bf PM} with $P^{(1p)}(D_{0}|\mathcal{T}_{\textrm{two\,red}}, \xi^{\textrm{PM}}) = p(2-p)$. This is less than or equal to the likelihood $P^{(1p)}(D_{0}|\mathcal{T}_{\textrm{RRRB}}, \xi^{\textrm{typO}}_{c = \{1,2,3\}}) = p(p^{2}-3p+3)$, for all $p$, which assumes a theory where only the first three cycles are red ($\mathcal{T}^{\star}=(1,1,1,0)\equiv\mathcal{T}_{RRRB}$). In addition, our reference framework's likelihood $P^{(1p)}(D_{0}|\mathcal{T}_{\textrm{two\,red}}, \xi^{\textrm{PM}})$ displays only parametric dominance over the likelihood associated with another framework not implementing {\bf PM}, namely ($\mathcal{T}_{\textrm{three\;red}}, \xi^{\textrm{typO}}_\mathcal{V}$), which takes the value $P^{(1p)}(D_{0}|\mathcal{T}_{\textrm{three\;red}}, \xi^{\textrm{typO}}_\mathcal{V})=\frac{3}{4}p(4-6p+4p^{2}-p^{3})$. Again, interestingly, {\bf PM} does not universally give rise to the highest likelihoods and in certain cases exhibits only a parameter-dependent dominance.

\section{Discussion}\label{SEC:Discussion}

The principle of mediocrity is a controversial issue in multiverse cosmology. According to one way of thinking, it articulates our intuitions about how one should reason from an appropriately defined, peaked probability distribution, to a prediction of a possible observation. But crucially, this intuition has been developed either in controlled laboratory settings, or more generally, in cases where we understand, and have some (at least theoretical) control over, the conditions that obtain in systems of interest. The multiverse, however, is a different story. 

It is plausible that we are \emph{more} typical of a set of appropriately restricted multiverse domains, but whether we can positively assert typicality heavily depends on who or what is predicted by the theories and the conditionalization schemes which we consider in multiverse cosmology. These latter schemes, at best, set down conditions that are necessary for the presence of `observers', but there is an ambiguity in defining precisely who or what these observers are. In addition we do not know precisely what parameters or conditions need to be fixed within the confines of any theory in order to unambiguously describe these observers. As a result, it is not clear that typicality is justified, even if we conditionalize in accord with the `ideal reference class' of~\citet{garriga+vilenkin_08}. Of course, we \emph{may} be typical, but following this line of thinking, we do not have good reason to assert that we are. 

The formalism of~\citet{srednicki+hartle_10} allows one to neatly address this multi-faceted concern, which affects our ability to reason in multiverse scenarios~\cite{smeenk_14}. Through their formalism, one can formulate a set of assumptions regarding typicality, and from this set, one can calculate relevant likelihoods for possible observations, to then see how well different assumptions do in terms of describing our observations. This is implemented in a Bayesian framework, so that \emph{if}, as we have assumed in this paper (see section~\ref{SEC:Bayesian}), we can assign equal amounts of prior credence to various candidate frameworks, then higher likelihoods translate to greater support for those frameworks given relevant experimental data. From the framework with the highest posterior probability then, one can \emph{infer} how typical we are. 

How well does typicality do? As we have discovered within the admittedly simplified multiverse setting of section~\ref{SEC:Multiverse_Model}: for a fixed theory, the principle of mediocrity yields likelihoods for our data that attain the supremum of all likelihoods considered, for all values of the probabilities of observers in domains.

But an important caveat is that this result is \emph{not} universal. Namely, if one is allowed to vary both the theory and the xerographic distribution implementing assumptions regarding typicality, the principle of mediocrity does not always provide the highest likelihoods. This is particularly pertinent when the set of candidate frameworks that constitute plausible alternatives for the description of a physical situation, do not always include xerographic distributions implementing the principle of mediocrity. 

One example where this occurs is in the situation where `Boltzmann brains' exist and out-number ordinary observers, but both sets of observers record the same data (see the discussion in~\cite{srednicki+hartle_10}, as well as~\cite{albrecht+sorbo_04, page_08, gott_08}). In this case, the first-person likelihood of our observations might be higher under the principle of mediocrity; but an unwanted consequence of favoring the corresponding framework, is the high likelihood of us \emph{being} Boltzmann brains. That is, this framework would also predict that our future measurements will be disordered, that is, uncorrelated with past measurements (as is presumed for Boltzmann brains). One way to avoid having to accept this consequence is by restricting the xerographic distribution accordingly---for example, by focussing attention on only a proper subset of appropriately chosen domains. This type of restriction has been actively developed in this paper, and is equivalent to an assumption of non-mediocrity under the scheme described in section~\ref{SEC:Xerographic_Distributions}.\footnote{One might wonder why we do not simply reject a theory that when partnered with the principle of mediocrity, does not constitute a plausible candidate framework. A response in the spirit of this paper, is that when the status of the principle of mediocrity is uncertain, it makes sense to examine the predictions of frameworks that constrain otherwise well-motivated theories by assumptions of non-mediocrity. A more pointed response is that one is simply not justified in rejecting a theory just because we would not be typical according to that theory. ~\citet{hartle+srednicki_07} raise this objection in discussing a (hypothetical) theory that predicts the existence of many more observers on Jupiter than on Earth. They claim it is unreasonable to reject such a theory, when we notice we are human and not Jovian, just because we would not be typical according to the theory.}

To sum up: if some of the frameworks one considers have xerographic distributions that do not implement the principle of mediocrity for relevant theories, then demanding that we favor a framework that includes the principle of mediocrity is hazardous. For as shown in section~\ref{SEC:PM_WBS}, we cannot guarantee it will produce the framework with the highest likelihood.

It is important to note that we have selected a particular reference class within which to implement the principle of mediocrity---namely, observers who witness our experimental data. The motivation for this choice was to test a limiting case of a `top-down' approach to conditionalization  (see~\cite{aguirre+tegmark_05, weinstein_06, garriga+vilenkin_08, azhar_14} for further details on top-down approaches). This limiting case requires consideration of the maximally specific reference class in the setup at hand. The assumption of typicality with respect to this reference class (encoded in $\xi^{\textrm{typD}}_{\mathcal{V}}$) then corresponds to the principle of mediocrity. A key result in this paper is that typicality with respect to this reference class does not necessarily give rise to the highest likelihoods for our data $D_{0}$, if one is allowed to vary both the theory and the xerographic distribution under consideration (as explained in section~\ref{SEC:PM_WBS}).

It is also important to note that we have evaluated frameworks based on first-person likelihoods generated for our data $D_{0}$. These computations are in accordance with the approach adopted by~\citet[\S IV]{srednicki+hartle_10}, who invoke such likelihoods in the evaluation of frameworks (where they also use xerographic distributions that correspond to each of $\xi^{\textrm{typO}}_{\mathcal{V}}$ and $\xi^{\textrm{typD}}_{\mathcal{V}}$). 

Another significant question to address, that manifestly goes beyond the cosmological settings studied in this paper, is: how should one evaluate the predictive power of frameworks (and in particular, the principle of mediocrity) in a case where one aims to predict the value of an observable, say, that is not explicitly included in the conditionalization scheme adopted? For realistic cosmological calculations, securing the required separation of the observable from the conditionalization scheme is non-trivial. One naturally requires that the observable being predicted is (i) correlated with the conditionalization scheme (otherwise the conditionalization scheme will play no role in the predictive framework), but (ii) is not perfectly correlated with the conditionalization scheme (otherwise one is open to the charge of circularity). Thus when it is not clear exactly how observables are correlated with the defining features of a conditionalization scheme, the need to strike a balance between (i) and (ii) gives rise to a difficult problem---namely, how to distinguish the observable to be predicted, from the defining features of the conditionalization scheme (see~\citet[\S III]{garriga+vilenkin_08} who mention such concerns).

Assuming that one has a solution to this problem, a formalism that can handle this type of predictive setting is described by~\citet[\S VI]{srednicki+hartle_10}. For the computation of first-person likelihoods, the appropriate way to proceed, as detailed by~\citet{garriga+vilenkin_08} and~\citet{hartle+hertog_13, hartle+hertog_15}, is to explicitly leave out that part of our data that involves the observable we aim to predict, in the specification of our conditionalization scheme. The assumption of typicality with respect to the reference class implicit in this specification, would then be the appropriate implementation of the principle of mediocrity~\cite{garriga+vilenkin_08}. 

It remains to apply the methods introduced by~\citet{srednicki+hartle_10}, and advanced in this paper, to more realistic cosmological settings, in order to more fully assess the extent of the errors that may arise from universally imposing the principle of mediocrity (see~\cite{hartle+hertog_13, hartle+hertog_15} for recent work in this direction). For now, our conclusion must be that the principle of mediocrity (in the style of~\cite{gott_93, vilenkin_95, page_96, bostrom_02, garriga+vilenkin_08}) is more questionable than has been claimed.

\begin{acknowledgments}
I am very grateful to Jeremy Butterfield for insightful discussions and comments on an earlier version of this paper. I thank audiences at Cambridge and at the Munich Center for Mathematical Philosophy at Ludwig-Maximilians-Universit{\"a}t M{\"u}nchen for helpful feedback. I am supported  by the Wittgenstein Studentship in Philosophy at Trinity College, Cambridge.
\end{acknowledgments}

\appendix*

\section{Calculation of $P^{(1p)}(D_{0}|\mathcal{T}, \xi^{\textrm{PM}})$ for the multiverse of cycles}

For the sake of completeness, we present the calculation for the first-person likelihood of our data $D_{0}$, given a theory $\mathcal{T}$, under the assumption of a xerographic distribution that imposes the principle of mediocrity, $\xi=\xi^{\textrm{PM}}:=\xi^{\textrm{typD}}_\mathcal{V}$. We work within the cyclic cosmological model of~\citet{srednicki+hartle_10} under the assumptions of section~\ref{SEC:SH_multiverse}; where we have $N$ domains stretched out in time, each with a probability $p$ of housing observers. From these assumptions, Eq.~(\ref{EQN:1stPersonD}) reduces to
\begin{widetext}
\begin{equation}
P^{(1p)}(D_{0}|\mathcal{T}, \xi^{\textrm{PM}}) =
  \sum_{\vec{\sigma}\in\mathcal{K}_{D_{0}}(\mathcal{V},\mathcal{T})} p^{\sum_{i=1}^{N}\sigma_{i}}(1-p)^{N-\sum_{j=1}^N\sigma_{j}}.
\label{EQN:1stPersonDHS}
\end{equation}
\end{widetext}

We organize the sum by separately considering configurations according to the total number of observers $n_{O}=\sum_{i=1}^{N}\sigma_{i}$ in each configuration $\vec{\sigma}$. Equation~(\ref{EQN:1stPersonDHS}) can then be thought of as $P^{(1p)}(D_{0}|\mathcal{T}, \xi^{\textrm{PM}}) = \sum_{n_{O}=1}^{N}G(p,n_{O}, N_{R})$, for some function $G$ we will compute shortly. The total number of red cycles $N_{R}$ depends on the theory $\mathcal{T}$, and we will separately consider the two cases into which the sum naturally partitions, namely,
$1 \leq n_{O} \leq  N_{R}$ and $N_{R} < n_{O} \leq  N$.

For $1 \leq n_{O} \leq  N_{R}$, we generate an expression for $G(p,n_{O}, N_{R})$ by sequentially placing all possible numbers of observers (out of a maximum of $n_{O}$ observers) in $N_{R}$ red cycles. In general, we find
\begin{widetext}
\begin{eqnarray}
G(p,n_{O}, N_{R})&=& p^{n_{O}}(1-p)^{N-n_{O}}\left[\binom{N_{R}}{n_{O}} +
 \binom{N_{R}}{n_{O}-1}\binom{N-N_{R}}{1} +\cdots+\binom{N_{R}}{1}\binom{N-N_{R}}{n_{O}-1}\right] \nonumber \\
&=& p^{n_{O}}(1-p)^{N-n_{O}}\sum_{k=0}^{n_{O}-1}\binom{N_{R}}{n_{O}-k}\binom{N-N_{R}}{k} \nonumber \\
&=& p^{n_{O}}(1-p)^{N-n_{O}}\left[\binom{N}{n_{O}}-\binom{N-N_{R}}{n_{O}}\right]\nonumber\\
&=& p^{n_{O}}(1-p)^{N-n_{O}}\sum_{k=1}^{N_{R}}\binom{N-k}{n_{O}-1},\nonumber
\end{eqnarray}
\end{widetext}
where we have used Vandermonde's identity in obtaining the third line, and have iterated using Pascal's formula in obtaining the fourth line. 

For $N_{R} < n_{O} \leq  N$, we obtain the same final result using a similar sequence of steps. Putting this all together, gives
\begin{widetext}
\begin{eqnarray}
P^{(1p)}(D_{0}|\mathcal{T}, \xi^{\textrm{PM}}) &=& \sum_{n_{O}=1}^{N}p^{n_{O}}(1-p)^{N-n_{O}}\sum_{k=1}^{N_{R}}\binom{N-k}{n_{O}-1}\nonumber\\
&=& \sum_{k=1}^{N_{R}}\sum_{n_{O}=1}^{N-k+1}\binom{N-k}{n_{O}-1}p^{n_{O}}(1-p)^{N-n_{O}}\nonumber\\
&=& \sum_{k=1}^{N_{R}}p(1-p)^{k-1}\sum_{m=0}^{N-k}\binom{N-k}{m}p^{m}(1-p)^{N-k-m}\nonumber\\
&=& p\sum_{k=1}^{N_{R}}(1-p)^{k-1}\nonumber\\
&=& 1-(1-p)^{N_{R}},
\end{eqnarray}
\end{widetext}
where we have used the binomial theorem to obtain the fourth line. This agrees with the corresponding expression (Eq. (5.5)) in~\cite{srednicki+hartle_10}, and Eq.~(\ref{EQN:FP_SH_Nr}) in the text above.

\bibliography{azhar_15_BIB_v2.bib}

\end{document}